\documentclass[twocolumn,showpacs,aps,prl,superscriptaddress]{revtex4}

\usepackage{graphicx}
\usepackage{dcolumn}
\usepackage{amsmath}
\usepackage{epsfig}

\input pubboard/babarsym

\newcommand{\BABARPubYear}    {04}
\newcommand{\BABARPubNumber}  {015}
\newcommand{\SLACPubNumber} {10462}

\def\figurebox#1#2#3{%
    \def\arg{#3}%
    \ifx\arg\empty
    {\hfill\vbox{\hsize#2\hrule\hbox to #2{\vrule\hfill\vbox to #1{\hsize#2\vfill}\vrule}\hrule}\hfill}%
    \else
    {\hfill\epsfbox{#3}\hfill}%
    \fi}

\def\Bflav {\ensuremath{B_{\text{flav}}}\xspace}
\def\Btag {\ensuremath{B_{\text{tag}}}\xspace}
\def\Bcp {\ensuremath{B_{CP}}\xspace}
\def\acp {\ensuremath{\mathcal A_{CP}}\xspace}

\def\KKKs  {\ensuremath{K^+K^-\KS}\xspace}
\def\KKKz  {\ensuremath{K^+K^-\Kz}\xspace}
\def\KKsKs {\ensuremath{K^+\KS\KS}\xspace}
\def\phiKs {\ensuremath{\phi \KS}\xspace}

\begin{document}

\preprint{\babar-PUB-\BABARPubYear/\BABARPubNumber} 
\preprint{SLAC-PUB-\SLACPubNumber}

\begin{flushleft}
%{\normalsize \babar\ Analysis Document \BADNumber, Version 16 }\\
\babar-PUB-\BABARPubYear/\BABARPubNumber \\
SLAC-PUB-\SLACPubNumber \\
%hep-ex/\LANLNumber\\[10mm]
%[10mm]
\end{flushleft}

%%%%%%%%%%%%%%%%%%%%%%%%%%%%%%%%%%%%%%%%%%%%%%%%%%%%%%%%%%%%%%%%%%%%%%%%%%%%%%%%%%
%									   	%%
%		              Title/Abstract					%%
%									   	%%
%%%%%%%%%%%%%%%%%%%%%%%%%%%%%%%%%%%%%%%%%%%%%%%%%%%%%%%%%%%%%%%%%%%%%%%%%%%%%%%%%%

\title{
{\large \bf
Branching Fractions and \emph{CP} Asymmetries 
in \boldmath{$\Bz \to \KKKs$\ } and \boldmath{$\Bp \to \KKsKs$} }}

% Author list
%% author list as of 02-Apr-2004 (603 authors)
%
\author{B.~Aubert}
\author{R.~Barate}
\author{D.~Boutigny}
\author{F.~Couderc}
\author{J.-M.~Gaillard}
\author{A.~Hicheur}
\author{Y.~Karyotakis}
\author{J.~P.~Lees}
\author{V.~Tisserand}
\author{A.~Zghiche}
\affiliation{Laboratoire de Physique des Particules, F-74941 Annecy-le-Vieux, France }
% ...
\collaboration{The \babar\ Collaboration}
\noaffiliation

\date{\today}% It is always \today, today, but you may specify any date with \date.

\begin{abstract}
We measure the branching fractions and \CP\ asymmetries in the decays 
$\Bz\to K^+ K^- \KS$\ and $\Bp \to K^+ \KS \KS$\ using a sample of approximately 122 million \BB\ pairs
collected by the \babar\ detector.
From a time-dependent analysis of the \KKKs\ sample that excludes \phiKs,\ the values of the \CP-violation parameters are 
$S = -0.56 \pm 0.25 \pm 0.04$
 and
$C = -0.10 \pm 0.19 \pm 0.10$,
where the first uncertainty is statistical, the second is systematic. We confirm that the final
state is nearly purely \CP -even. Using this result and setting $C=0$, we extract the Standard Model parameter 
$\sin{2\beta} = 0.57 \pm 0.26 \pm 0.04 ^{+0.17}_{-0}$
where the last error is due to uncertainty on the \CP\ content.
We present the first measurement of the \CP-violating charge asymmetry 
	$\acp (B^+ \to \KKsKs) = -0.04 \pm 0.11 \pm 0.02$, 
with a 90\% confidence-level interval of $\left [-0.23, 0.15 \right ]$.
The branching fractions are
${\mathcal B}(B^0 \to K^+K^-\Kz) = (23.8 \pm 2.0 \pm 1.6) \times 10^{-6}$ and 
${\mathcal B}(B^+ \to \KKsKs) = (10.7 \pm 1.2 \pm 1.0) \times 10^{-6}$.
\end{abstract}

% PACS, the Physics and Astronomy Classification Scheme.

\maketitle

%%%%%%%%%%%%%%%%%%%%%%%%%%%%%%%%%%%%%%%%%%%%%%%%%%%%%%%%%%%%%%%%%%%%%%%%%%%%%%%%%%
%									   	%%
%		              Introduction					%%
%									   	%%
%%%%%%%%%%%%%%%%%%%%%%%%%%%%%%%%%%%%%%%%%%%%%%%%%%%%%%%%%%%%%%%%%%%%%%%%%%%%%%%%%%

In the Standard Model (SM) of particle physics, the decays $\Bz \to K^+K^-\KS$\ and $\Bp \to K^+\KS\KS$~\cite{charge} 
are dominated by $b\rightarrow s\bar{s}s$ gluonic penguin diagrams~\cite{sPenguin}. 
\CP violation in such decays arises from the Cabibbo--Kobayashi--Maskawa (CKM) quark-mixing mechanism~\cite{ckm}.
Neglecting CKM-suppressed contributions, the expectation for the \CP-asymmetry parameters in 
$\Bz \to \Kp\Km\KS$\ decays is the same as in $\Bz\to\jpsi\KS$\ decays,
where \CP\ violation has been observed~\cite{Aubert:2002ic, Abe:2003yu}.
The decay rates for $\Bp \to \Kp\KS\KS$\ and $\Bm \to \Km\KS\KS$\ are expected to be equal.
However, contributions from physics beyond the SM could invalidate these predictions~\cite{Grossman:1996ke}.
Since $b\rightarrow s\bar{s}s$ decays involve one-loop transitions, 
they are especially sensitive to additional contributions.
Present results in decays of neutral $B$ mesons are inconclusive due to large statistical errors.
Belle measures the \CP\ asymmetry parameter in \phiKs\ decays of
$\sin2\beta = -0.96 \pm 0.50 ^{+0.09}_{-0.11}$~\cite{Abe:2003yt} which is 3.5 standard deviations from the
SM expectation
of $\sin{2\beta}=0.731 \pm 0.056$~\cite{Aubert:2002ic, Abe:2003yu}.
A \babar\ measurement of $\sin2\beta = 0.47 \pm 0.34 ^{+0.08}_{-0.06}$~\cite{Aubert:2004ii}
is consistent with the SM and disagrees with Belle by 2.3 standard deviations.

A more accurate \CP\ measurement can be made using all the decays to
\KKKs\ that do not contain a $\phi$ meson.
This sample is several times larger than the sample of \phiKs~\cite{Aubert:2003hz, Garmash:2003er}.
As Belle noted~\cite{Garmash:2003er}, the \CP\ content of the final state can be extracted using an isospin analysis.
In decays that exclude \phiKs,\ Belle measures
$\sin{2\beta}=0.51 \pm 0.26 \pm 0.05 ^{+0.18}_{-0}$~\cite{Abe:2003yt}, 
consistent with the SM expectation.
In this letter we present measurements of \CP\ asymmetry and \CP\ content in \KKKs\ decays,
and the first measurement of the charge asymmetry rate in $\Bp \to \Kp\KS\KS$\ decays.

%%%%%%%%%%%%%%%%%%%%%%%%%%%%%%%%%%%%%%%%%%%%%%%%%%%%%%%%%%%%%%%%%%%%%%%%%%%%%%%%%%
%									   	%%
%		           Event  Selection   					%%
%									   	%%
%%%%%%%%%%%%%%%%%%%%%%%%%%%%%%%%%%%%%%%%%%%%%%%%%%%%%%%%%%%%%%%%%%%%%%%%%%%%%%%%%%

This analysis is based on about 122 million \BB\ pairs collected
with the \babar\ detector~\cite{Aubert:2001tu} at the \pep2\
asymmetric-energy \epem\ storage rings at SLAC, operating on the $\Upsilon(4S)$
resonance.
We reconstruct $B$ mesons from $\KS\to\pi^+\pi^-$ and $K^\pm$ candidates.
Charged kaons are distinguished from pions and protons using energy-loss (\dedx) information in the tracking
system and from the Cherenkov angle and number of photons measured by 
the detector of internally reflected Cherenkov light (DIRC).
We accept $\KS\rightarrow\pi^+\pi^-$ candidates that have a two-pion invariant mass within $12$~\mevcc\ of the nominal \KS\ mass~\cite{Hagiwara:fs}, 
a decay length greater than 3 standard deviations, 
and a cosine of the angle between the line connecting the $B$ and \KS\ decay vertices and the \KS\ momentum greater than 0.999.
The three daughters in the $B$ decay are fitted constraining their paths to a common vertex, and the \KS\ mass to
the nominal value.

In the characterization of the $B$ candidates we use two kinematic variables.
The energy difference $\Delta E = E_B - \sqrt{s}/2$ is reconstructed from the energy of the $B$ candidate $E_B$ 
and the total energy $\sqrt{s}$ in the \epem\ center-of-mass (CM) frame.
The  \DeltaE\  resolution for signal events is 18~\mev.
We also use the beam-energy-substituted mass $\mes = \sqrt{({s}/{2} + {\vec{p}}_{i} \cdot  {\vec{p}}_{B} )^{2}/{E^{2}_{i}} - {{\vec{p}}_{B}}^{\,2}}$,
where $( {\vec{p}}_{i}, E_{i} )$ is the four-momentum of the initial \epem\ system and 
${\vec{p}}_{B}$ is the momentum of the $B$ candidate, both measured in the laboratory frame.
The \mes\ resolution for signal events is 2.6~\mevcc.\ 
We retain candidates with  $|\DeltaE|<200$~\mev and $5.2<\mes<5.3$~\gevcc.

The background is dominated by random combinations of tracks created in $\epem\to q\bar{q}~(q=u,d,s,c)$ continuum events. 
We suppress this background by utilizing the difference in the topology in the CM frame between jet-like $q\bar{q}$ events 
and spherical signal events.
The topology is described using angle $\theta_T$ between the thrust axis of the $B$ candidate 
and the thrust axis of the charged and neutral particles in the rest of the event~(ROE)~\cite{Aubert:2001tu}.
Other quantities that characterize the event topology are two sums over the ROE:  $L_0=\sum|\vec{p_i}^*|$ and $L_2=\sum |\vec{p_i}^*| \cos{^2\theta_i}$,
where $\theta_i$ is the angle between the momentum $\vec{p_i}^*$ and the thrust axis of the $B$ candidate.
Additional separation is achieved using the angle $\theta_B$ between the $B$-momentum direction and the beam axis.
After requiring $|\cos\theta_T| < 0.9$, these four event shape variables are combined into a Fisher discriminant $\mathcal F$~\cite{Fisher:et}.

The remaining background originates from $B$ decays where a neutral or charged pion is missed during
reconstruction (peaking $B$ background).
We use Monte Carlo (MC) events to model the signal and the peaking background,
and data sidebands to model continuum background.

We suppress background from $B$ decays that proceed through a $b \to c$ transition leading to the \KKKs (\KKsKs)\ final state 
by applying invariant mass cuts to remove $\Dz \to KK$, $\Dp \to \Kp\KS$,
$\jpsi \to KK$, and $\chi_{c0} \to KK(\KS\KS)$ decays.
Finally, $B$ decays into final states with pions are eliminated by requiring
the pion misidentification rate to be less than 2\%.

%%%%%%%%%%%%%%%%%%%%%%%%%%%%%%%%%%%%%%%%%%%%%%%%%%%%%%%%%%%%%%%%%%%%%%%%%%%%%%%%%%
%									   	%%
%		           CP  Measurement 					%%
%									   	%%
%%%%%%%%%%%%%%%%%%%%%%%%%%%%%%%%%%%%%%%%%%%%%%%%%%%%%%%%%%%%%%%%%%%%%%%%%%%%%%%%%%

The time-dependent \CP asymmetry is obtained by measuring the proper time difference \deltat\ between
a fully reconstructed neutral $B$ meson (\Bcp) decaying into \KKKs, and the partially reconstructed 
recoil $B$ meson (\Btag). Decay products of the recoil side are used to determine the \Btag\ meson's flavor (flavor tag) and
to classify the event into five mutually exclusive tagging categories~\cite{Aubert:2002ic}. 
If the fraction of events in category $c$ is $\epsilon_c$
and the mistag probability is $\mistag_c$, the overall quality of the tagging, $\sum_c \epsilon_c (1-2w_c)^2$, is ($28.0 \pm 0.4$)\%.

The time difference \deltat\  is extracted from the measurement of the separation \deltaz\ between the \Bcp\ and \Btag\ vertices, 
along the boost axis ($z$) of the \BB\ system.
The vertex position of the \Bcp\ meson is reconstructed primarily from kaon tracks,
and its MC-estimated resolution ranges between 40--80\mum, depending on the opening angle and direction of the kaon pair.
The final \deltat\ resolution is dominated by the uncertainty on the \Btag\ vertex
which allows the \deltat~(\deltaz) precision with r.m.s. of 1.1~ps~(180~\mum).
We retain events that have $|\deltat|<20$~ps and whose estimated uncertainty $\sigma_{\deltat}$ is less than 2.5~ps.
The \deltat\ resolution function is parameterized as a sum of two Gaussian distributions
whose widths are given by a scale factor times the event-by-event uncertainty $\sigma_{\deltat}$. 
A third Gaussian distribution, with a fixed large width, accounts for a small
fraction of outlying events~\cite{Aubert:2002ic}.

Parameters describing the tagging performance and the \deltat\ resolution function  are extracted from 
approximately 30,000 \Bz\ decays into $D^{(*)-}X^+\,(X^+ = \pip, \rho^+, a_1^+)$ 
flavor eigenstates (\Bflav\ sample). 

The decay rate ${\text{f}}_+({\text{f}}_-)$ when the flavor of the tagging meson is a \Bz~(\Bzb)\ is given by 
\begin{eqnarray}
{\text{f}}_\pm(\, \deltat)& = &{\frac{{\text{e}}^{{- \left| \deltat 
\right|}/\tau_{\Bz} }}{4\tau_{\Bz}}}  \, [
\ 1 \hbox to 0cm{}
\pm 
S \sin{( \deltamd  \deltat )}  \nonumber \\
& & 
\mp 
\,C  \cos{( \deltamd  \deltat) }   ],
\label{eq::timedist}
\end{eqnarray}
where $\tau_{\Bz}$ is the mean \Bz\ lifetime and \deltamd\ is the \Bz--\Bzb oscillation frequency.
The parameters $C$ and $S$ describe the magnitude of \CP violation in the decay and
the interference between decay and mixing, respectively.
In the SM, we expect $C=0$ because there can be no direct \CP\ violation when there is only one decay mechanism.
If we exclude \phiKs\ events by applying a $K^+K^-$ invariant mass cut of 15~\mevcc\  around the nominal $\phi$ mass~\cite{Hagiwara:fs},
and assume that the remaining \Bcp\ candidates are \CP-even, as our analysis below indicates,
we expect $S=-\sin{2\beta}=-0.731 \pm 0.056$~\cite{Aubert:2002ic, Abe:2003yu}.

Direct \CP violation in $\Bp\to \KKsKs$\ decays is measured as an asymmetry in the decay rates
\begin{equation}
\acp = \frac{\Gamma_{K^-\KS\KS} - \Gamma_{K^+\KS\KS}}{\Gamma_{K^-\KS\KS} + \Gamma_{K^+\KS\KS}}\,.
\end{equation}
The SM expectation for \acp\ is zero.

%%%%%%%%%%%%%%%%%%%%%%%%%%%%%%%%%%%%%%%%%%%%%%%%%%%%%%%%%%%%%%%%%%%%%%%%%%%%%%%%%%
%									   	%%
%		           Likelihood and Results 				%%
%									   	%%
%%%%%%%%%%%%%%%%%%%%%%%%%%%%%%%%%%%%%%%%%%%%%%%%%%%%%%%%%%%%%%%%%%%%%%%%%%%%%%%%%%

Branching fractions and \CP\ asymmetries are extracted in unbinned extended maximum likelihood fits to the different samples.
The likelihood function ${\mathcal L}$, with event yields $N_i$ and 
probability density functions (PDFs) ${\mathcal P}_{i,j}$, is: 
\begin{equation}
{\mathcal L} = \exp{\left(-\sum_{i}N_{i}\right)}
\prod_{j=1}\left[\sum_{i}N_{i}{\mathcal P}_{i,j}\right]
\end{equation}
where $j$ runs over events and $i$ over event yields.
We have a total of 6144 events in the \KKsKs\ mode, and 13864~(12862) in the \KKKs\ mode with
\phiKs\ included~(excluded).

\begin{figure}[hb]
\center
\vspace*{-3.8mm}
\epsfig{file=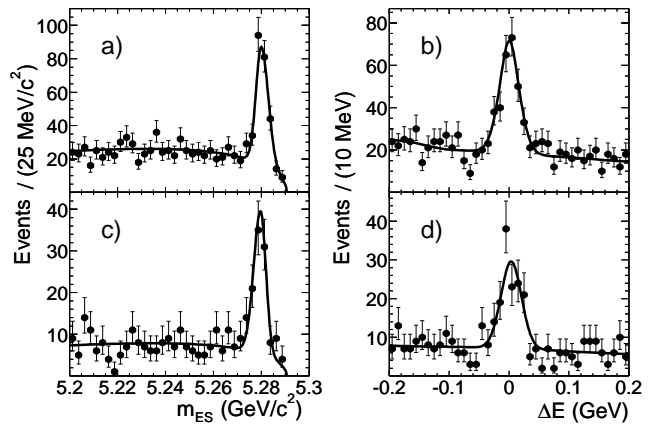, width=8.5cm}
\caption{Projection plots of the variables \mes (a, c) 
and \DeltaE (b, d) in the fits
for $\Bz\to\KKKs$\ (top) and $\Bp \to \KKsKs$\ (bottom) decays.
The points are data and the curves are projections from the likelihood fit. 
The signal-to-background ratio is enhanced with a cut on the event probability.
\label{fg::yields}}
\end{figure}

In the measurement of the branching fractions \BR,\  the total PDF is formed as  
${\mathcal P}(\mes) \cdot {\mathcal P}(\DeltaE) \cdot {\mathcal P}({\cal F})$.
Event yields for signal, continuum, and peaking $B$ background are varied in the fit.
In the extraction of the charge asymmetry \acp\ in \KKsKs\ decays, the yields are split by the charge, which
brings the total number of varied parameters to six.
To extract the branching fractions, we assign a weight for each event to belong to the signal decay, 
${\mathcal W}_j=\frac{ \sum_i V_{s,i} {\mathcal P}_{i,j} }{ \sum_i N_i {\mathcal P}_{i,j} }$
where $V_{s,i}$ is the signal row of the covariance matrix obtained from the fit~\cite{sPlot}.
The branching fraction is calculated as $\BR=\sum_j {\mathcal W}_j/\varepsilon_j$.  
Since the efficiency $\varepsilon_j$ varies across the phase space, $\varepsilon_j$
is computed in small phase-space bins using simulated events. 
The method is cross-checked with a simple counting analysis. 
Distributions of \mes\ and \DeltaE\ are shown in Fig.~\ref{fg::yields} and
the fit results are given in Table~\ref{tb::results}.

\begingroup
\begin{table*}[btp]
\caption{\label{tb::results}Summary of branching fraction (\BR), time-dependent ($S$, $C$) and direct \CP-asymmetry (\acp) results.
$N_{\text{sig}}$ and $\eps$ are the signal yield and the  average total efficiency in the branching-fraction fit; 
$f_{even}$ is the \CP-even  fraction of the final states. The 90\% confidence-level interval for \acp\ is $[-0.23, 0.15]$.
}
\begin{ruledtabular}
\setlength{\extrarowheight}{1.5pt}
\footnotesize
\begin{tabular}{lccccccc}
Mode   	  & $\eps$ (\%) & $N_{\text{sig}}$ &     \BR\ ($10^{-6}$)   &$f_{even}$ 	&  $S$         		    &        $C$       	        & \acp \\
\hline
\KKKz$^{\CP}$& $8.58$    & $201 \pm 16$ 	   & $20.2 \pm 1.9 \pm 1.4$ &$0.98 \pm 0.15 \pm 0.04$	& $-0.56 \pm 0.25 \pm 0.04$   
& $-0.10 \pm 0.19 \pm 0.10$ & --- \\
\KKKz$^{\mbox{all}}$  & $8.78$      & $249 \pm 20$     & $23.8 \pm 2.0 \pm 1.6$ &$0.83 \pm 0.12 \pm 0.03$	&      ---                   &  ---                      & --- \\
\hline
\KKsKs    & $9.7$        & $122 \pm 14$     & $10.7 \pm 1.2 \pm 1.0$ &     ---	       &   $-0.16\pm 0.35$          & $-0.08\pm 0.22$  &  $-0.04 \pm 0.11 \pm 0.02$ \\
\end{tabular}
\setlength{\extrarowheight}{0pt}
\end{ruledtabular}
\raggedright
{$^{\CP}$Excludes \phiKs\ events.}
\end{table*}
\endgroup

In the time-dependent \CP fit, \KKKs\ events that exclude \phiKs\ decays are fit simultaneously with the \Bflav\ sample.
The PDFs are formed as   
${\mathcal P}(\mes) \cdot {\mathcal P}(\DeltaE) \cdot {\mathcal P}({\cal F}) \cdot {\mathcal P}_c(\deltat;\sigma_{\deltat})$
for \Bcp events and 
${\mathcal P}(\mes) \cdot {\mathcal P}_c(\deltat;\sigma_{\deltat})$
in the \Bflav\ sample. 
The \deltat\ resolution and tagging parameters are allowed to be different for each tagging category $c$.
Fit parameters that are common to both samples are the signal fractions in tagging categories $\epsilon_c$,
the average mistag fraction $\mistag_c$, 
the difference between \Bz\ and \Bzb\ mistag rates $\Delta\mistag_c$, 
and the \deltat\ resolution functions for signal 
and background 
events.
We also vary the \KKKs\ signal yield and background yields in tag categories, 
the \CP\ parameters, 
and the parameters describing the \deltat\ shape of the background. 
The total number of floated parameters is 38.
The largest correlation between $S$ or $C$ with any linear combination of 
other parameters is 6.6\%.

\vspace{1mm}
\begin{figure}[hb]
\begin{center}
\begin{tabular}{c}
\epsfig{file=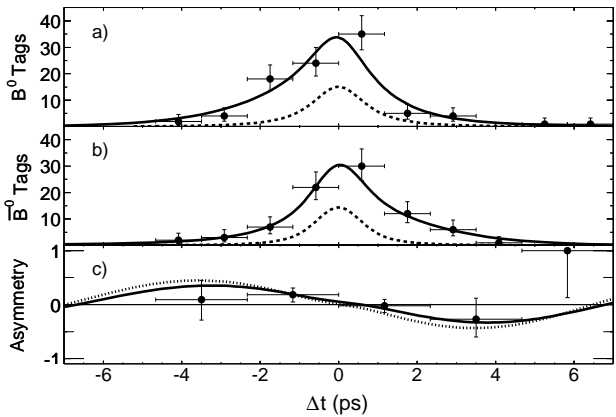, width=8.8cm}
\end{tabular}
\vspace*{-1.3cm}
\caption{Plots a) and b) show the $\Delta t$ distributions of  
\Bz- and \Bzb-tagged \KKKs\ events. 
The solid lines refer to the fit for all events; the dashed lines correspond to the background. 
Plot c) shows the raw asymmetry, where the solid line is obtained from the fit
and the dotted line is the SM expectation for the measured \CP\ content.
The signal-to-background ratio is enhanced with a cut on the event probability.
\label{fg::dt}}
\vspace*{-0.4cm}
\end{center}
\end{figure}

Results of the time-dependent \CP\-asymmetry measurement in \KKKs\ are  given in  Table~\ref{tb::results}.
Figure~\ref{fg::dt} shows the \deltat\ distributions of events with \Bz\ and \Bzb\ tags, with projections from the likelihood fit superimposed.
The fit procedure is verified with the \KKsKs\ sample (Table~\ref{tb::results}), where one expects zero asymmetry,
and the $\jpsi\KS$\ sample where the results are consistent with our previous measurement~\cite{Aubert:2002ic}.

%%%%%%%%%%%%%%%%%%%%%%%%%%%%%%%%%%%%%%%%%%%%%%%%%%%%%%%%%%%%%%%%%%%%%%%%%%%%%%%%%%
%									   	%%
%		           CP even fraction					%%
%									   	%%
%%%%%%%%%%%%%%%%%%%%%%%%%%%%%%%%%%%%%%%%%%%%%%%%%%%%%%%%%%%%%%%%%%%%%%%%%%%%%%%%%%

We evaluate the fraction $f_{\text{even}}$ of \CP-even final states in $\Bz\to\KKKs$\ decays by comparing 
\KKKz\ and \KKsKs\ decay rates: 
$
	f_{\text{even}}=\frac{ 2 \Gamma(\Bp \to \KKsKs)}{\Gamma(\Bz \to \Kp\Km\Kz)}
$
~\cite{Garmash:2003er}. The results 
listed in Table~\ref{tb::results} are in agreement with Belle's measurements of $0.86 \pm 0.15 \pm 0.05$ and $1.04 \pm 0.19 \pm 0.06$
for  the total sample and the \CP\ sample that excludes \phiKs\ events, respectively~\cite{Garmash:2003er}.
We estimate the fraction of remaining \phiKs\ events in the \CP\ sample, using a non-interfering Breit-Wigner for the $\phi$ shape
and measured branching fractions, to be $1.1\pm0.4$\%.
As a consistency check, we examine the distribution of the cosine of the helicity angle 
$\theta_H$, which is defined as the angle between the $K^+$ and \Bz\ directions in the $\Kp\Km$ center of mass frame. 
The  distribution in several $\Kp\Km$ invariant mass bins of the \CP\ sample is approximately uniform which is consistent with S-wave decays.
The presence of interference effects due to \CP -odd amplitudes cannot be ruled out, but this
would require a full amplitude analysis which is not feasible with the present statistics.

If we account for a small \CP -odd fraction in the \CP\ sample, we can extract the SM parameter $\sin 2\beta$.
In a fit with $C=0$ we get  $\sin{2\beta} = -S/(2f_{even}-1) = 0.57 \pm 0.26 \pm 0.04 ^{+0.17}_{-0}$
where the last error is due to uncertainty on the \CP\ content.

%%%%%%%%%%%%%%%%%%%%%%%%%%%%%%%%%%%%%%%%%%%%%%%%%%%%%%%%%%%%%%%%%%%%%%%%%%%%%%%%%%
%									   	%%
%		           Systematics     					%%
%									   	%%
%%%%%%%%%%%%%%%%%%%%%%%%%%%%%%%%%%%%%%%%%%%%%%%%%%%%%%%%%%%%%%%%%%%%%%%%%%%%%%%%%%

\begin{table}[th]
\caption{Branching fraction systematic uncertainties\,(\%).
\label{tb::systBF}}
\begin{tabular}{lcc}
\hline \hline
Source				&	\KKKs		&	\KKsKs		\\
\hline
Efficiency	 		&	5.6		&	8.6		\\
PDF parameterization		&	2.7		&	2.5		\\
Non-charm \BB\ background	&	2.2		&	2.9		\\
Charm \BB\ background		&	1.2		&	1.0		\\
Other				&	1.7		&	1.6		\\
\hline
Total				&	6.9		&	9.6		\\
\hline \hline
\end{tabular}
\end{table}
Systematic uncertainties in the branching fraction measurements are given in Table~\ref{tb::systBF}.
We include contributions from the signal reconstruction efficiency and from the modeling of the efficiency variation over the phase space.
Other errors come from the fit bias, the counting of \BB\ pairs, and the misidentification of kaons.
We assume equal production rates of \BzBzb and \BpBm.
The systematic uncertainty on \acp\ due to charge asymmetry in track finding and identification is 0.02.

\begin{table}[ht]
\caption{Systematic uncertainties in \CP\ parameters.
\label{tb::systCP}}
\begin{tabular}{lcc}
\hline \hline
Source				&	$S$	&	$C$	\\
\hline
Fit bias			&	0.024	&	0.026	\\
DCSD				&	0.018	&	0.053	\\
Detector effects		&	0.013	&	0.012	\\
Tag asymmetries			&	0.010	&	0.078	\\
Other				&	0.016	&	0.012	\\
\hline
Total				&	0.04	&	0.10	\\
\hline \hline
\end{tabular}
\end{table}
The systematic errors on the time-dependent \CP -asymmetry parameters are given in Table~\ref{tb::systCP}.
The errors account for  the fit bias, the presence of double CKM-suppressed decays(DCSD) in \Btag~\cite{Long:2003wq}, 
uncertainty in the beam spot and detector alignment, and
the  asymmetry in the tagging efficiency for signal and background events.
Other smaller effects come from \deltat\ resolution, PDF parameterization of yield variables, and uncertainty 
on the \Bz\ lifetime and mixing frequency.
In the fit we use $\tau_{\Bz}=1.537\pm0.015$~ps and $\deltamd =0.502\pm0.007~{\rm \ps}^{-1}$~\cite{Hagiwara:fs}.

%%%%%%%%%%%%%%%%%%%%%%%%%%%%%%%%%%%%%%%%%%%%%%%%%%%%%%%%%%%%%%%%%%%%%%%%%%%%%%%%%%
%									   	%%
%		           Summary         					%%
%									   	%%
%%%%%%%%%%%%%%%%%%%%%%%%%%%%%%%%%%%%%%%%%%%%%%%%%%%%%%%%%%%%%%%%%%%%%%%%%%%%%%%%%%

In summary, we have measured branching fractions for charmless decays of $B$ mesons into the three-body
final states $\Bz \to \KKKz$\ and $\Bp \to \KKsKs$.
Using two independent approaches, we find that the \KKKs\ final state is dominated by a \CP -even component.
The results agree with previous measurements~\cite{Aubert:2003hz,Garmash:2003er}.
In the first measurement of the charge asymmetry in $\Bp \to \KKsKs$\ decays, we find no evidence for direct \CP\ violation.
We measure a time-dependent \CP\ asymmetry in $\Bz\ \to \KKKs$ decays at the $1.9\sigma$ level.
The obtained $\sin{2\beta}$ is consistent with the SM expectation and previous measurements in decays into the
\KKKs\ final state~\cite{Abe:2003yt,Aubert:2004ii}, 
but differs from Belle's measurement in \phiKs\ decays~\cite{Abe:2003yt} by 2.7 standard deviations.

%%%%%%%%%%%%%%%%%%%%%%%%%%%%%%%%%%%%%%%%%%%%%%%%%%%%%%%%%%%%%%%%%%%%%%%%%%%%%%%%%%
%									   	%%
%		            Acknowledgements        				%%
%									   	%%
%%%%%%%%%%%%%%%%%%%%%%%%%%%%%%%%%%%%%%%%%%%%%%%%%%%%%%%%%%%%%%%%%%%%%%%%%%%%%%%%%%
We are grateful for the excellent luminosity and machine conditions
provided by our \pep2\ colleagues, 
and for the substantial dedicated effort from
the computing organizations that support \babar.
The collaborating institutions wish to thank 
SLAC for its support and kind hospitality. 
This work is supported by
DOE
and NSF (USA),
NSERC (Canada),
IHEP (China),
CEA and
CNRS-IN2P3
(France),
BMBF and DFG
(Germany),
INFN (Italy),
FOM (The Netherlands),
NFR (Norway),
MIST (Russia), and
PPARC (United Kingdom). 
Individuals have received support from the 
A.~P.~Sloan Foundation, 
Research Corporation,
and Alexander von Humboldt Foundation.

%%%%%%%%%%%%%%%%%%%%%%%%%%%%%%%%%%%%%%%%%%%%%%%%%%%%%%%%%%%%%%%%%%%%%%%%%%%%%%%%%%
%									   	%%
%		            Bibliography        				%%
%									   	%%
%%%%%%%%%%%%%%%%%%%%%%%%%%%%%%%%%%%%%%%%%%%%%%%%%%%%%%%%%%%%%%%%%%%%%%%%%%%%%%%%%%


\begin{thebibliography}{99}
\bibitem{charge}
Charge-conjugate states are included unless explicitly stated otherwise.
\end{thebibliography}
\end{document}